\documentclass[prb,showpacs,twocolumn]{revtex4}
\usepackage{epsfig}
\newcommand{\Sigm}{{\mit\Sigma}}
\newcommand{\ab}[1]{{\rm #1}}
\newcommand{\upd}{\ab{d}}
\begin{document}
\bibliographystyle{revtex}
\title{Ladder approximation to spin velocities in quantum wires}
\author{Lars Kecke}
\affiliation{Physikalisches Institut, Universit\"at Freiburg,
Hermann-Herder-Str.\ 3, D-79104 Freiburg, Germany}
\author{Wolfgang H\"ausler}
\affiliation{Physikalisches Institut, Universit\"at Freiburg,
Hermann-Herder-Str.\ 3, D-79104 Freiburg, Germany}
\affiliation{I.\ Institut f\"ur Theoretische Physik, Universit\"at Hamburg,
Jungiusstr.\ 9, D-20355 Hamburg, Germany}

\begin{abstract}
The spin sector of charge-spin separated single mode quantum
wires is studied, accounting for realistic microscopic
electron-electron interactions. We utilize the ladder
approximation (LA) to the interaction vertex and exploit
thermodynamic relations to obtain spin velocities. Down to not
too small carrier densities our results compare well with
existing quantum Monte-Carlo (QMC) data. Analyzing second order
diagrams we identify logarithmically divergent contributions as
crucial which the LA includes but which are missed, for example,
by the self-consistent Hartree-Fock approximation. Contrary to
other approximations the LA yields a non-trivial spin
conductance. Its considerably smaller computational effort
compared to numerically exact methods, such as the QMC method,
enables us to study overall dependences on interaction
parameters. We identify the short distance part of the
interaction to govern spin sector properties.
\end{abstract}
\maketitle
At low energies the one-dimensional (1D) electron liquid
exhibits charge-spin separation \cite{solyom,MdC}, in contrast
to the Fermi liquids of higher dimensionality \cite{anderson}.
Theoretical considerations based on the Tomonaga-Luttinger
liquid (TLL) \cite{Tomo} have predicted this behavior
\cite{solyom} and recent experiment \cite{Goni,Auslaender} has
provided strong evidence supporting different velocities
$v_{\rho}$ and $v_{\sigma}$. Both velocities differ from the
Fermi velocity $v_\ab{F}$: in the Hubbard model, for example,
charge density waves (plasmons) propagate faster,
$v_{\rho}>v_\ab{F}$, while spin density waves propagate slower,
$v_{\sigma}<v_\ab{F}$, than the Fermi velocity for repulsive
on-site interaction \cite{schulz90}.

Quantitative knowledge of $v_{\sigma}$ is
decisive to predict spin transport \cite{balents,rapid} and magnetic
properties \cite{Voit}. Within the random phase approximation
(RPA), or when treating left and right going particles as distinguishable
\cite{DL,MdC} as originally done by Luttinger\cite{Tomo},
$v_\sigma=v_\ab{F}$ stays unrenormalized with interactions
\cite{overhauser}. On the other hand, even in first order perturbation
theory the exchange or Fock term, proportional to the $2k_F$-Fourier
component of the electron-electron interaction 
$V(q)$, influences the spin velocity.
(In contemporary quantum wires the range of the interaction
usually exceeds the Fermi wavelength, so $V(q=0)\neq V(q=2k_\ab{F})$).
Quantitatively, however,
this latter approach is limited to $V(2k_\ab{F})\ll v_\ab{F}$
(corresponding to electron gas parameters $r_\ab{s}\ll 1$).
Furthermore, it spoils SU(2) spin rotation invariance of the
microscopic electron model, see below. Quantum Monte Carlo
(QMC) has demonstrated that $v_{\sigma}/v_\ab{F}$ decreases
with increasing inter-particle repulsion in quantum wires
\cite{creffield}, in qualitative resemblance to the Hubbard
model \cite{schulz90}. Values of $v_{\sigma}/v_\ab{F}\approx 0.5$
have been estimated \cite{creffield} for present day single channel
quantum wires \cite{tarucha,rother,yacoby}.

Numerically exact techniques, however, are computationally
extremely demanding, especially in spin sector. This yet has
prohibited scans through larger parts of the parameter space
that characterizes the microscopic interaction; in Ref.\
\onlinecite{creffield} only one interaction range and one
channel width has been investigated. A future study of
multi-channel quantum wires would require using considerably
bigger systems. Here, a sufficiently accurate and tractable
approximative scheme would be helpful. Among the techniques
established for Fermi liquids the Hartree-Fock approximation
(HF) has proven as useful to estimate boundary exponents
\cite{schonhammer} and, when carried out self-consistently, to
yield amazingly quantitative results in charge sector
\cite{whahm}; it captures for example very well the
non-monotonous dependence of $K_{\rho}$ on the electron density
\cite{whlkahm}, beyond the RPA. On the other hand, the mean
field approximation has turned out to fail badly in spin sector
\cite{whlkahm}. Below we show that in a perturbative language
this failure can be traced back to the wrong class of diagrams
summed by the self consistent HF. In the present work we sum up
ladder diagrams and demonstrate that they comprise a
`complementary' class of diagrams that account much better for
spin properties.

The ladder approximation (LA) to the effective interaction
vertex originally has been established to study strongly
correlated fermion systems with short range interactions, such
as nuclear matter \cite{brueckner} but it performs remarkably well
for the the also strongly correlated charge sector of a
1DES\cite{Nafari,Agosti}.

In comparison to
other Fermi-liquid methods \cite{Agosti}, such as the Singwi, Tosi,
Land, and S\"olander (STLS) approximation scheme 
\cite{nagano,nagano2,yang}, the LA is known to account well for
short distance properties of the interaction when tested\cite{nagano}
with exactly solvable
models \cite{yang}. As we shall demonstrate, this short distance
behavior is indeed most relevant to magnetic properties.
In this paper we generalize the LA to allow for non-zero
magnetizations and for spin currents to gain homogeneous and
static spin susceptibilities. Summing ladder
diagrams yields estimates to $v_{\sigma}$ that compare well with
existing QMC--data at not too low electron densities while the
numerical effort stays comparable to self consistent HF
calculations. This enables us to scan different parameters of
the microscopic interaction. Of prime interest, {\em e.g.}\ to
carbon nanotubes are the interaction range and the diameter of
the quantum wires.

\section{Model}
To model the microscopic interaction we use \cite{whlkahm}
\begin{equation}\label{interaction}
V(q)=\frac{2v_\ab{F}}{k_\ab{F}a_\ab{B}}\left[\ab{K}_0(qd)-
\ab{K}_0(q\sqrt{d^2+4R^2})\right]
\end{equation}
in momentum space,
where $\ab{K}_0$ denotes a modified Bessel function and $a_\ab{B}$
is the Bohr radius. This form accounts for the
experimentally important parameters, the interaction range $R$,
given by the distance to the nearest metal gate, and the diameter $d$
of the wire, which determine $V(q)$ at small and at large $q$, respectively.
Details of the wire's cross section affect the interaction only
at $q>d^{-1}$, {\em i.e.}\ at carrier densities where occupation
of higher subbands starts. We do not consider this case here.

In Fermion representation the quantum wire (length $L$, $s=\pm
1$ denotes spin) is described by
\begin{eqnarray}
H&=&\sum_{k,s}\epsilon(k)c_{k,s}^\dagger c_{k,s}^{}+\nonumber\\
{}&&+\frac{1}{2L}\sum_{k,s,k',s',q}c_{k-q,s}^\dagger c_{k'+q,s'}^\dagger
V(q)c_{k',s'}^{}c_{k,s}^{}\;,\label{H}
\end{eqnarray}
$c_{k,s}$ are Fermi operators in the wave number basis.
For non-linear single particle dispersion $\epsilon(k)=k^2/2m$,
as in semiconducting quantum wires of effective carrier mass
$m$ and in the presence of $2k_\ab{F}$-scattering between
antiparallel spins near opposite Fermi-points bosonization of
model (\ref{H}) introduces terms of higher than quadratic order
in the Bosonic density fluctuations. One could attempt, at
least in principle, to eliminate higher order terms by the RG
method in course of which they would renormalize all pre-factors
of the quadratic terms, {\em i.e.}\ the TLL velocities
$v_{N\nu}$, and $v_{J\nu}$ in
\begin{equation}\label{Htll}
H_\ab{TLL}=\sum_{\nu=\sigma,\rho}v_{\nu}\sum_{q\ne 0}h_{\nu,q}+
\frac{\pi}{4L}\sum_{\nu=\sigma,\rho}(v_{N\nu}N_\nu^2+v_{J\nu}J_\nu^2)\;.
\end{equation}
In practice such a RG approach to the quantitative values of the
TLL parameters has not been tested and seems not promising for
strong interactions. On the other hand, there is little doubt
that the one-dimensional electron liquid remains in the TLL phase
even at strongest interactions.
Fortunately, exact thermodynamic relations allow
quantitative determination of the TLL velocities $v_{N\nu}$ and
$v_{J\nu}$, related with the total charge or spin $N_{\nu}$ and
the total currents $J_{\nu}$, from homogeneous and static
susceptibilities since the latter are observable quantities.

Symmetries reduce the number of independent TLL-parameters to be
determined for given microscopic single particle dispersion
$\epsilon(k)$ and interaction $V(q)$. Left-right symmetry,
$\epsilon(k)=\epsilon(-k)$ leads to the TLL-relations
\begin{equation}\label{vrel}
v_\nu^2=v_{N\nu}v_{J\nu}
\end{equation}
for the sound velocities $v_{\nu}$ at which Bosonic charge or
spin density fluctuations move, as described by the operators
$h_{\nu,q}$ in (\ref{Htll}). Most quantum wires, furthermore,
show Galilei invariance in charge sector which leads to
$v_{J\rho}=v_\ab{F}$, independent of the interaction. Spin
rotation SU(2) invariance, present in (\ref{H}) without a Zeeman
field or spin-orbit coupling \cite{rapid}, ensures the three
velocities $v_{\sigma}=v_{N\sigma}=v_{J\sigma}$ to be equal in
spin sector \cite{Voit}.

Here we concentrate on spin sector, where the thermodynamic
relations read (see {\em e.g.}\ Ref.\ \onlinecite{Voit})
\begin{eqnarray}
v_{N\sigma}&=&\frac{2}{\pi\chi}=
\frac{2L\partial^2E_0}{\pi\partial N_\sigma^2}=
\frac{\pi}{2}\frac{\partial^2E_0/L}{\partial P^2}\nonumber\\
\mbox{and}\quad v_{J\sigma}&=&\frac{2L\partial^2E_0}
{\pi\partial J_\sigma^2}=
\frac{\pi}{2}\frac{\partial^2E_0/L}{\partial I^2}\;.\label{thermo}
\end{eqnarray}
$P=\pi N_\sigma/2L$ and $I=\pi J_\sigma/2L$ represent
magnetization and spin current, respectively.

For later comparison with the Hartree-Fock approximation
it is useful to express ground state energies and susceptibilities
in terms of the self-energy
\begin{equation}\label{selfenergy}
\Sigm_s(k,\omega)={G^0_s}(k,\omega)^{-1}-G_s(k,\omega)^{-1}\;.
\end{equation}
$G_s(k,\omega)$ is the full electron propagator with
respect to (\ref{H}) and $\:G^0_s(k,\omega)=\left(\omega-k^2/2m+
\ab{sign}[(k-sI)^2-(k_\ab{F}+sP)^2]i0\right)^{-1}\:$ the
free propagator. The ground state energy, required in
(\ref{thermo}), can be expressed as \cite{FW}
\begin{equation}
E_0/L=\sum_{s}\frac{1}{4\pi^2}\int \upd k \upd\omega
\left(\frac{k^2}{2m}+\omega\right)G_s(k,\omega)
\end{equation}
which reduces to
\begin{equation}\label{esig}
E_0/L=\sum_{s}\frac{1}{2\pi}\int\limits^{k_\ab{F}+s(I+P)}_
{-k_\ab{F}+s(I-P)}
\upd k\left[\frac{k^2}{2m}+\frac{1}{2}\Sigm_s(k)\right]
\end{equation}
for self energies not depending on frequency $\omega$
(and therefore are real as a consequence of the Lehmann
representation of $G_s$). This condition is fulfilled within
the HF ({\em cf.\/} Eq.~(\ref{hfr}) below) and within the LA.
We are not aiming to determine temporal or spatial
correlation functions in the Fermionic representation, but
rely here on the exact solution \cite{DL,haldane} of the Boson
model (\ref{Htll}).

It can serve as a test to the quality of any approximative
scheme in spin sector whether or not it respects the condition
\cite{Voit} $v_{N\sigma}=v_{J\sigma}$ dictated by SU(2)
invariance. The HF, for example, leaves $v_{J\sigma}=v_\ab{F}$
independent of interactions since it ignores
$2k_\ab{F}$--interactions between electrons of opposite spins
and thus violates spin rotation symmetry. The STLS
only allows to calculate magnetic susceptibilities
$v_{N\sigma}$ in spin sector \cite{CG}, so it cannot be tested
in its behavior regarding the SU(2) symmetry. The LA finally
does indeed renormalize $v_{J\sigma}$, though only by about
half the amount it renormalizes the value for $v_{N\sigma}$
compared to $v_\ab{F}$. In so far we find that the LA does not
fully obey the SU(2) symmetry but proves as superior to the other
approximative schemes.

\section{Ladder Approximation}
To introduce the LA we follow the References~\onlinecite{FW}
and \onlinecite{Nafari} but generalize the calculations for
finite magnetization $P$ and spin current $I$. The self-energy
shift
\begin{eqnarray}
\Sigm_s(k;P,I,k_\ab{F})&=&\frac{1}{2\pi}\sum_{s'}
\int\limits^{k_\ab{F}+s'(I+P)}_{-k_\ab{F}+s'(I-P)}\!\!\!\!\!\!\!\!
\upd k' \times\label{sigg}\\
&&\hspace*{-3ex}\biggl(g_{ss'}(k,k',0)-\delta_{ss'}g_{ss'}(k,k',k-k')
\biggr)\nonumber
\end{eqnarray}
due to interactions with the sea of other electrons is expressed
in terms of an effective (Brueckner) interaction matrix
$g_{ss'}(k,k',q)$. Knowing $g_{ss'}(k,k',q)$ exactly would
yield the exact ground state energy through (\ref{esig}) and, by
virtue of (\ref{thermo}), the exact values for the
TLL-velocities. The Brueckner interaction matrix is
closely related with the static structure factor which often is
exploited to investigate how the short range part of the
interaction affects short distance correlations of
one-dimensional electron liquids \cite{Nafari,calmels}.

For the ladder approximation a Bethe--Salpeter integral equation
\begin{eqnarray}
g_{ss'}(k,k',q)&=&V(q)+\frac{1}{2\pi}\int\limits_{-\infty}^\infty\upd p\;
g_{ss'}(k,k',p)\nonumber\\
&&\times K_{ss'}(k,k',p)V(q-p)\label{bse}
\end{eqnarray}
has to be solved, describing multiple scattering between
electrons that otherwise propagate freely outside
the Fermi sphere according to
\begin{eqnarray}
&&K_{ss'}(k,k',p)=\\
&&\displaystyle 2m\frac{\Theta(|k-sI-p|-k_{\ab{F}s})
\Theta(|k'-s'I+p|-k_{\ab{F}s'})}{k^2+k'^2-(k-p)^2-(k'+p)^2}\;.
\nonumber\label{bsekern}
\end{eqnarray}
In (\ref{bsekern}) we have defined spin dependent Fermi-momenta
$k_{\ab{F}s}\equiv k_\ab{F}+sP$ at finite magnetizations. We
solve equation~(\ref{bse}) for each pair of momenta $(k,k')$ and
spins $(s,s')$ the Householder method after mapping the infinite
$p$ range to $[0,\pi]$ by $p=k_F\cot\phi$ and discretizing
$\phi_i=\pi(i-1)/(N-1)$. A value of $N=150$ turned out as an
optimal compromise between CPU time (increasing with $N^3$) and
accuracy (the error in $g\sim N^{-2}$ due to the cusps
of $g(k,k',q)$ at $q=2k_\ab{F}$). Only at strong interactions,
when spin velocities drop below half of the Fermi velocity, we
used $N=250$. Finally, ground state energies are integrated via
(\ref{esig}) and (\ref{sigg}) using the trapezoid rule on a grid
of 129 points over the interval $[-2k_\ab{F};2k_\ab{F}]$ which
sufficed to accurately resolve the effects of small
magnetizations.

\begin{figure}
\centerline{\epsfig{file=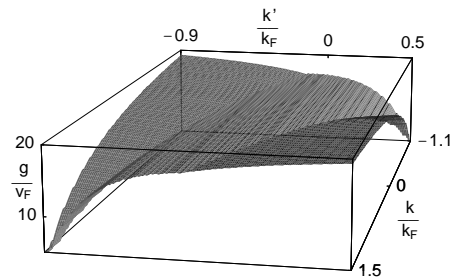,width=6cm}}
\caption{Self-consistent $g_{\uparrow\downarrow}(k,k',0)$
for $R/d=14.14$, $k_\ab{F}d=0.15$, $P=0.3\,,\,I=0.2$. Due to
correlation effects the Brueckner interaction vertex shows
pronounced valleys along $k+k'=\pm 2P$.
The bare value $V(0)=22.29v_\ab{F}$.}\label{g0}
\end{figure}

The solution $g_{ss'}(k,k',q)$ exhibits pronounced dependence on
$k$ and $k'$, {\em cf.}\ Figure~\ref{g0}, arising primarily from
the short distance correlations at $k=-k'$. Ignoring this
dependence, as it is commonly done in 3D ({\em cf.}\ Ref.\
\onlinecite{calmels} and references therein), would clearly not
be justified. This correlation has similarities to the striking
anti-ferromagnetic $2k_\ab{F}$ modulations found in the
self-consistent HF--density in 1D \cite{whlkahm} which also
cannot be ignored for reliable results in the charge sector.

\section{Results}
\begin{figure}
\centerline{\epsfig{file=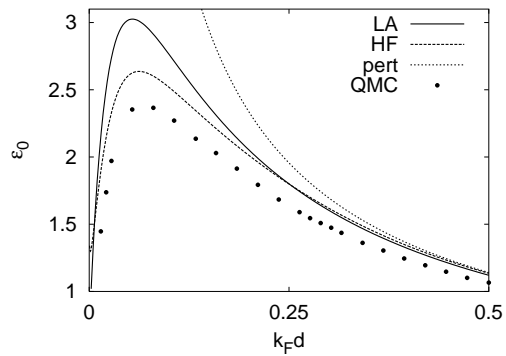, angle=-90,width=7cm}}
\caption{Ground state energy $\varepsilon_0=E_0/(Lk_\ab{F}^3/m)$
versus $k_\ab{F}d$ for $R/d=14.14$. The electron density
$2k_{\rm F}/\pi$ can be expressed through the electron gas
parameter $r_\ab{s}=\pi/(8k_{\rm F}a_\ab{B})$ at $d=a_\ab{B}/2$.
`LA' are our results, `pert' denotes first order perturbation
theory when $g(k,k',q)\equiv V(q)$. `HF' refers to self-consistent
Hartree--Fock results of Ref.\ \onlinecite{whlkahm} and the
QMC data are taken from Ref.\ \onlinecite{creffield}.}\label{efig}
\end{figure}

\begin{figure}
\centerline{\epsfig{file=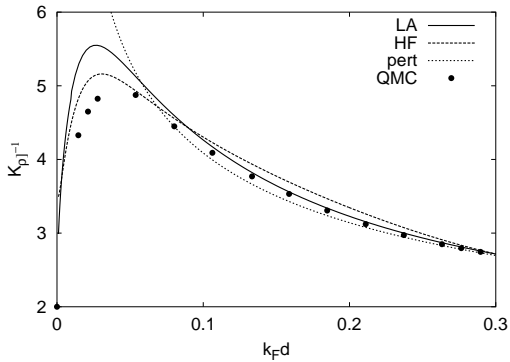, angle=270, width=7cm}}
\caption{Charge sector Luttinger parameter $K_\rho$ versus $k_\ab{F}d$
for $R/d=14.14$. Labels as in Figure~\ref{efig}.}\label{kfig}
\end{figure}

\begin{figure}
\centerline{\epsfig{file=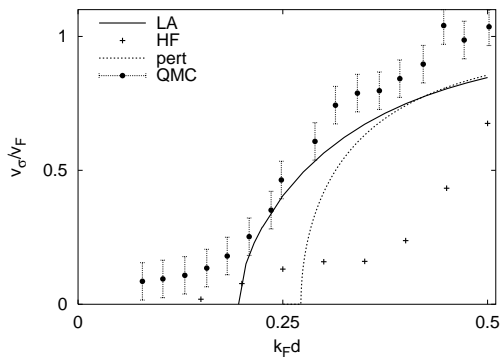, angle=-90,width=7cm}}
\caption{Spin velocities in units of $v_\ab{F}$ versus
$k_\ab{F}d$ for $R/d=14.14$. The HF data are taken from
Ref.\ \onlinecite{whlkahm} and the QMC data from Ref.\
\onlinecite{creffield}.}\label{vfig}
\end{figure}

\begin{figure}
\centerline{\epsfig{file=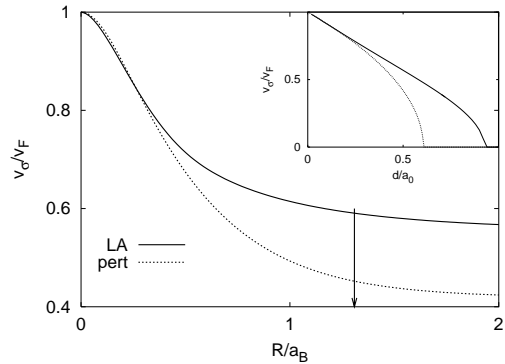, angle=-90,width=7cm}}
\caption{Spin velocity versus interaction range $R$ for
$k_\ab{F}d=0.3$. At the arrow $R=\pi/4k_\ab{F}=2r_\ab{s}a_\ab{B}$.
When $R\ll d$ the interaction (\ref{interaction}) vanishes
trivially; we always assume $R\gg d$. The inset shows spin velocity
versus wire width for $k_\ab{F}d=0.3$ at $R/d=14.14$.}\label{vRfig}
\end{figure}

Figure~\ref{efig} shows ground state energies (in units of the
respective kinetic energies) versus carrier density using
different approximations. QMC data \cite{creffield} can be
regarded as exact within symbol size. Remarkably, at densities
$\:k_\ab{F}d\gtrsim 0.2\:$, where self-consistence is seen in
Figure~\ref{efig} to still improve the HF--estimate to $E_0$
compared to `pert' (and then provides the optimum Fermi--liquid
wave function), the LA yields slightly lower ground state
energies. As seen in Figure~\ref{kfig} the LA also yields better
values of $K_\rho$ in this regime. Only at smaller electron
densities HF-theory engenders lower ground state energies
and a better estimate to the $K_\rho$-parameter ($k_\ab{F}d<0.1$).
This energy gain is accompanied by spontaneous symmetry breaking
and pronounced (though unreal) static $4k_\ab{F}$--periodic Wigner
crystal-like modulations \cite{whlkahm} of the charge density
in the HF ground state. However, this success of the HF in charge
sector does clearly not carry over to the spin sector \cite{whlkahm}
as seen in Figure~\ref{vfig}.

Remarkably, in the LA the spin velocity
follows the QMC-data down to rather low electron densities,
superior to other electron gas theory approaches.
Only at small densities $v_{\sigma}^{\ab{LA}}$
vanishes, pretending the transition into a ferromagnetic ground
state which is not expected to occur in 1D for finite range
interactions. The LA does not reproduce the behavior
$v_\sigma\propto k_\ab{F}^2$ derived from the Hubbard model at
small fillings \cite{coll} to which quantum wires should cross
over \cite{whlkahm} when the inter-particle spacing exceeds the
interaction range, $k_\ab{F}R\ll \pi/2$.

Figure~\ref{vRfig} demonstrates that at interaction ranges
$k_\ab{F}R\gg\pi/2$ the value of $R$ does not affect the spin
velocity, unlike the ground state energy or the charge sector
exponent which both depend logarithmically on $R/d$
\cite{gogolin,whlkahm}. This is consistent with the
perturbative result according to which $V(k=2k_\ab{F})$ but not
$V(k=0)$ governs the magnetic susceptibility where the former
$V(2k_\ab{F})\sim 2v_\ab{F}\ab{K}_0
(2k_\ab{F}d)/k_\ab{F}a_\ab{B}$ becomes independent of
$k_\ab{F}R\gg 1$. This is of particular relevance to carbon
nanotubes \cite{spintransportintubes} where $k_\ab{F}R$ can be
of the order of $10^3$. The inset of Figure~\ref{vRfig}
complements Figure~\ref{vfig}, showing how the spin velocity
varies with wire width $d$ at fixed $k_\ab{F}d=0.3$.

\section{Discussion}
To analyze further why the LA captures magnetic properties so
well we compare it diagrammatically with the self-consistent
HF. For this purpose we use the self-energy (\ref{selfenergy})
were HF and LA can be compared directly. We expand the
self-energy (\ref{sigg}) to second order in the interaction.
The first order contribution, obtained by putting
$g(k,k',q)=V(q)$ in (\ref{sigg}), yields
\begin{equation}\label{storder}
\Sigm_{s,\ab{pert}}(k+sI)=\frac{4k_\ab{F}}{2\pi}V(0)-
\int\limits_{-k_\ab{F}-sP}^{k_\ab{F}+sP}\upd k'\;\frac{V(k-k')}{2\pi}\;.
\end{equation}
The exchange term on the right hand side of (\ref{storder}) is
effective only for parallel spins and therefore independent of
the spin current $I$ so that $v_{J\sigma}$ remains
un-renormalized to this order.

Self-consistent HF is described by
\begin{eqnarray}
\Sigm_{s,\ab{HF}}(k)&=&\frac{-i}{4\pi^2}\sum_{s'}\int\upd k'\upd\omega\;
\times\nonumber\\
&&(V(0)-\delta_{ss'}V(k-k'))G_{s'}(k',\omega)\label{hfr}\\
G_s(k,\omega)&=&G^0_s(k,\omega)+G_s^0(k,\omega)\Sigm_{s,\ab{HF}}(k)
G_s(k,\omega)\;.\label{hfg}
\end{eqnarray}
To first order, $\Sigm_{s,\ab{HF}}(k)$ agrees with
(\ref{storder}). We see that (\ref{hfr}) will not depend on
$I$, even when self-consistence $G^0\to G$ is reached.

\begin{figure*}
\centerline{\epsfig{file=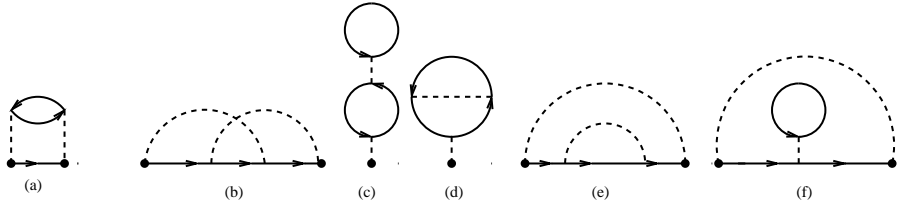, angle=-90,width=14cm}}
\caption{Irreducible electron self-energy $\Sigm(k)$ in
second order perturbation theory. The LA accounts for the
contributions $(a)$ and $(b)$ while the HF includes the terms
$(c)-(e)$. At $k=k_{\ab{F}s}$ the term $(a)$ behaves like
$P\ln P$ yielding a divergent contribution to
$v_\sigma$.}\label{div}
\end{figure*}

Second order contributions are obtained after iterating
(\ref{sigg}) and (\ref{bse}) for the LA or (\ref{hfr}) and
(\ref{hfg}) for the HF, respectively. For the LA this yields
\begin{widetext}
\begin{eqnarray}\label{lasigm}
\lefteqn{\Sigm^{(2)}_{s,\ab{LA}}(k)-\Sigm_{s,\ab{pert}}(k)=}\\
(a)&+&\frac{1}{(2\pi)^4}\sum_{s'}\int \upd k'\upd\omega\, G^0_s(k',\omega)
\int \upd p \upd\omega'\, V(p)^2 G^0_{s'}(k-p,k^2/2m-\omega')
G^0_{s'}(k'+p,k'^2/2m+\omega')\nonumber\\
(b)&-&\frac{1}{(2\pi)^4}\int \upd k'\upd\omega\, G^0_s(k',\omega)
\int \upd p \upd\omega'\, V(p)V(k-k'-p)G^0_s(k-p,k^2/2m-\omega')
G^0_s(k'+p,k'^2/2m+\omega')\nonumber\\
\lefteqn{\hspace{-10mm}\raisebox{0mm}[5mm][3mm]
{\mbox{while for the HF}}}\nonumber\\
\lefteqn{\Sigm^{(2)}_{s,\ab{HF}}(k)-\Sigm_{s,\ab{pert}}(k)=}
\label{hfsigm}\\
(c)&-&\frac{1}{(2\pi)^4}V(0)^2\sum_{s',s''}\int \upd k'\upd\omega\,
G^0_{s'}(k',\omega)\int \upd p \upd\omega'\, G^0_{s''}(p,\omega')^2
\nonumber\\
(d)&+&\frac{1}{(2\pi)^4}V(0)\sum_{s'}\int \upd k'\upd\omega\,
G^0_{s'}(k',\omega)\int \upd p \upd\omega'\,
V(k'-p)G^0_{s'}(p,\omega')^2\nonumber\\
(e)&-&\frac{1}{(2\pi)^4}\int \upd k'\upd\omega\, G^0_s(k',\omega)
\int \upd p \upd\omega'\, V(k-p)V(p-k')G^0_s(p,\omega')^2\nonumber\\
(f)&+&\frac{1}{(2\pi)^4}V(0)\sum_{s'}\int \upd k'\upd\omega\,
G^0_{s'}(k',\omega)\int \upd p \upd\omega'\,
V(k-p)G^0_s(p,\omega')^2\;.\nonumber
\end{eqnarray}
\end{widetext}
These are all irreducible second order contributions. None of
the diagrams, depicted in Figure~\ref{div}, occurs in both of the
approximations, insofar the LA and the HF can be regarded as
complementary. It is known \cite{DL} that any finite (but the
first) order self-energy contribution diverges logarithmically
for $k\to k_\ab{F}$; to second order this occurs only due to
the `LA terms' $(a)$ and $(b)$ while the HF contributions
remain finite.

Furthermore, only term $(a)$ breaks Galilei invariance in spin
sector as required to satisfy the SU(2) symmetry condition
according to which the spin conductance should be equally
renormalized by interactions as the magnetic susceptibility,
$v_{J\sigma}=v_{N\sigma}$. Term $(a)$ in (\ref{lasigm})
describes scattering between antiparallel $s=-s'$ spins at
opposite Fermi points which affects
$\partial_I\Sigm_s^{(2)}(k;P,I,k_\ab{F})$ and the spin
conductance $v_{J\sigma}$.

In order to compare the importance of the different terms $(a)$
to $(f)$ regarding $\partial_P\Sigm_s^{(2)}(k;P,I,k_\ab{F})$ and
thus the magnetic susceptibility $v_{N\sigma}$ we consider for
the moment a contact interaction, $V(q)=V$. Then all second order
contributions can be calculated analytically. The sum of all
spin parallel parts $s=s'$ cancel (spin parallel Fermions do not
interact at contact) and only the $s=-s'$ part of $(a)$ and 
the term $(c)$ (with $s'=s$ or $s'=s''$) remain non-vanishing. The
latter is not divergent while term $(a)$ can be expressed by
sums of dilogarithms \cite{dilog} which can further be analyzed
for $I,P\ll k_\ab{F}$. The result is
\begin{eqnarray}
(a)&\propto& mV^2[\pi^2/3-(|P+I|)/k_F\ln((|P+I|)/k_F)-\nonumber\\
&&2(P+I)/k_F+O(P^2,I^2)]\label{sigmcont}
\end{eqnarray}
at $k=k_\ab{F}+sP+sI$ and $I\leftrightarrow -I$ in
(\ref{sigmcont}) near the other Fermi point $k=-k_\ab{F}-sP+sI$.
The infinite slope seen in equation (\ref{sigmcont}) at $I=P=0$
(by virtue of (\ref{thermo}) and (\ref{esig})) results in
logarithmically diverging second order contributions to
$\:v_{J\sigma}-v_\ab{F}\:$ and to $\:v_{N\sigma}-v_\ab{F}\:$ as
a function of $P$ or $I$ which dominate over the non-diverging
HF-term $(c)$. The LA comprises in summing the leading
logarithmically diverging contributions to the spin velocity
which is one reason for its success regarding spin sector
properties. A finite interaction range does not remove
logarithmic divergencies to qualitatively alter this observation.

\section{Outlook}
In conclusion, we have generalized the ladder approximation (LA)
to investigate the spin sector of single channel quantum wires in
the presence of a realistic microscopic interaction. While the
numerical effort is considerably smaller, we obtain values for
the spin velocities that compare well with existing quantum
Monte-Carlo data at not too small particle densities. The LA
accounts for interaction diagrams which renormalize the spin
conductance. Furthermore, the LA diagrams include the leading
logarithmically diverging contributions to spin conductance and
susceptibility and thus are of dominant importance for magnetic
properties. The self-consistent Hartree-Fock approximation
(HF) misses these diagrams and therefore leaves the spin
conductance unaffected by interactions leading to an erroneous
result for the spin velocity. The short distance part of the
interaction is identified to govern spin sector properties.
Metal gates fabricated close to a quantum wire will screen the
long range part of the interaction but leave spin properties
almost unaffected.

The LA should prove useful to study spin properties of
multi-channel quantum wires where the numerically exact methods
would be even more demanding than already in the single channel
case. Compared to the HF the LA does not suffer from
incommensurate densities $k_{\ab{F}\uparrow}\ne k_{\ab{F}
\downarrow}$ at finite magnetizations. This advantage over the
HF might even carry over to the charge sector of coupled
quantum channels since it is likely that incommensurate particle
densities in different channels will cause similar instabilities
of the HF-procedure as for magnetic properties in single
channels \cite{whlkahm}. Thus, the LA might prove as the method
of choice also for the charge sector of multi-channel quantum
wires.
\acknowledgments
We thank H.\ Grabert for valuable discussions and C.E.\ Creffield
for useful communications on the QMC results.

\bibliography{lk007}
\end{document}